\documentclass[a4paper,11pt,usenames,dvipsnames]{article}
\pdfoutput=1 

\usepackage{jheppub}
\usepackage[T1]{fontenc} 
\usepackage{amsmath,amssymb,amsfonts,amsthm}
\usepackage{graphicx}
\usepackage[usenames,dvipsnames]{color}
\usepackage{enumitem}
\usepackage{verbatim}
\usepackage[percent]{overpic}
\usepackage{rotating}
\usepackage{hyperref}
\usepackage{array}
\usepackage{mathtools}
\usepackage{lmodern}
\usepackage[normalem]{ulem}
\usepackage{braket}
\usepackage{tensor}
\usepackage{dsfont}
\usepackage{bm}
\usepackage{tikz}

\usepackage{mathrsfs}

\usetikzlibrary{decorations.pathmorphing}
\usepackage{CJKutf8}

\usetikzlibrary{positioning}
\usetikzlibrary{calc,through,backgrounds}

\theoremstyle{definition}
\newtheorem{definition}{Definition}[section]
\newtheorem{theorem}{Theorem}[section]

\newcommand{\kako}[1]{\left( #1 \right)}
\newcommand{\kagikako}[1]{\left[ #1 \right]}
\newcommand{\ts}[1]{ _{\text{#1}} }

\newcommand{\erfc}{\text{erfc}}

\allowdisplaybreaks

\DeclareMathOperator{\Tr}{Tr}
\newcommand{\R}{\mathbb{R}}
\newcommand{\dd}{\text{d}}
\newcommand{\bk}{{\bm{k}}}
\newcommand{\bx}{{\bm{x}}}

\newcommand{\id}{\mathds{1}}
\newcommand{\sx}{\mathsf{x}}

\newcommand{\ii}{\mathsf{i}}

\newcommand{\kk}{|\bm{k}|}

\title{Quantum Otto engine driven by quantum fields}

\author[a]{Kensuke Gallock-Yoshimura,}
\author[a]{Vaishant Thakur,}
\author[a]{and Robert B. Mann}

\affiliation[a]{Department of Physics and Astronomy, University of Waterloo, Waterloo, Ontario, N2L 3G1, Canada}

\emailAdd{kgallock@uwaterloo.ca}
\emailAdd{v2thakur@uwaterloo.ca} 
\emailAdd{rbmann@uwaterloo.ca}

\abstract{We consider a quantum Otto engine using an Unruh-DeWitt particle detector model which interacts with a quantum scalar field in curved spacetime. 
We express a generic condition for extracting positive work in terms of the effective temperature of the detector. 
This condition reduces to the well-known positive work condition in the literature under the circumstances where the detector reaches thermal equilibrium with the field. 
We then evaluate the amount of work extracted by the detector in two scenarios: an inertial detector in a thermal bath and a circulating detector in the Minkowski vacuum, which is inspired by the Unruh quantum Otto engine. 
}

\begin{document} 
\maketitle
\flushbottom

\section{Introduction} \label{sec:introduction}

Quantum thermodynamics is a study of thermodynamic phenomena from the perspective of quantum mechanics. 
Within this field quantum heat engines, where the working substance is quantum matter that interacts with thermal baths, have emerged as a crucial area of research \cite{Scovil.heat.engine.maser}. 
Notable among quantum heat engines are the quantum Carnot engine and the quantum Otto engine (QOE). 
The quantum Carnot engine operates through two isothermal processes and two quantum adiabatic processes. 
The QOE, on the other hand, is composed of two isochoric thermalization processes and two quantum adiabatic processes \cite{Scovil.heat.engine.maser, Feldmann.QOE.2000, Kieu.secondlaw.demon.otto, Kieu.heat.engine.2006, Rostovtsev.Otto.2003, Quan.multilevel.engine.2005, Quan.multilevel.heat.engine}. 
For a comprehensive review, see \cite{Myers.quantumThermo.review}.

Our interest lies in the QOE using a two-level quantum system in the realm of relativistic quantum information (RQI), where relativity is taken into account. 
In particular, we consider a specific model for a qubit known as the Unruh-DeWitt (UDW) particle detector model \cite{Unruh1979evaporation, DeWitt1979}, which interacts with a quantum field in (curved) spacetime. 
The UDW detector model exhibits the essential features of the light-atom interaction as long as angular momenta are not exchanged \cite{PhysRevD.87.064038, PhysRevA.89.033835}. 
A UDW detector reveals interesting phenomena such as the Unruh effect \cite{Unruh1979evaporation}, where a uniformly accelerating UDW detector perceives a thermal bath at the Unruh temperature $T\ts{U}=a/2\pi$ (where $a$ is the magnitude of the acceleration), whilst an observer at rest sees a vacuum.

Investigations have been carried out previously on the QOE employing the UDW detector model. 
In \cite{UnruhOttoEngine}, it was shown that a linearly accelerating UDW detector can be utilized to extract work in a QOE. 
This process is facilitated by the Unruh effect, which generates hot and cold thermal baths by varying the magnitude of acceleration $a$. 
Such an engine is commonly referred to as the Unruh QOE. 
The amount of work extracted is determined by a response function of the UDW detector, which is the probability of a transition from the ground state to the excited state, and vice versa. 
Subsequent studies have further expanded on the Unruh QOE in various scenarios, including a fermionic field \cite{Finn.UnruhOtto}, a degenerate detector \cite{Xu.UnruhOtto.degenerate}, entangled detectors \cite{Kane.entangled.Unruh.Otto, Barman.entangled.UnruhOtto}, and in the presence of a reflecting boundary \cite{Mukherjee.UnruhOtto.boundary}. 
However, these studies focus on a specific acceleration trajectory, the state of a field, and a switching function. 
A comprehensive investigation of the QOE in a more general setting still remains largely unexplored.

In this paper, we explore the QOE within the framework of RQI in a generalized setting. 
We consider a pointlike UDW detector following an arbitrary trajectory in curved spacetime, interacting with a quantum scalar field in a quasi-free state. 
Thus, the thermalization of the detector is not necessarily required. 
By employing a perturbative method, we derive general expressions \eqref{eq:free p work extracted} and \eqref{eq:extracted work wrt response} for the work extracted through the quantum Otto cycle, which are written in terms of the response function of the detector. 
In addition, we identify conditions for extracting positive work in a generalized setting, given by \eqref{eq:free p positive work condition} and \eqref{eq:positive work condition wrt effective temp}. 
These conditions are expressed in terms of the effective temperature, an estimator for temperature, as perceived by the detector. 
Since thermality is not a prerequisite, the expressions for work extracted and the positive work condition are applicable to any situation. 
Notably, when the field is in a thermal state, known as the Kubo-Martin-Schwinger (KMS) state, and the interaction duration of a rapidly decreasing switching function is sufficiently long, the effective temperature becomes the KMS temperature. 
In such a case, the condition for positive work \eqref{eq:positive work condition wrt effective temp} becomes identical to the original condition found in \cite{Feldmann.QOE.2000, Kieu.secondlaw.demon.otto}.

The derived expressions are demonstrated concretely within two scenarios: an inertial detector immersed in a KMS state of a quantum field in Minkowski spacetime and a detector in circular motion in the Minkowski vacuum. 
The first example is the most basic instance of thermality within QFT, for which we employ a Gaussian switching function and explore the effects of interaction duration.

The second example is inspired by the Unruh QOE previously examined in the literature, which has some challenges. 
A linearly accelerating detector requires an immense amount of acceleration to reach a temperature of 1 Kelvin. 
This can be seen from the expression of the Unruh temperature in SI units, $T\ts{U}=\hbar a/2\pi k\ts{B}c$, where $\hbar, k\ts{B}$, and $c$ are respectively the reduced Planck constant, the Boltzmann constant, and the speed of light. 
It is well known that an acceleration of $a\sim 10^{20}$ m/s$^2$ is required to achieve $T\ts{U}\sim 1$ K. 
This implies that significant work must be performed on the detector to extract a small amount of work from the Unruh QOE. 
Moreover, the duration of interaction is almost instantaneous. 
Typically, the Unruh QOE assumes that the detector initially travels at a constant speed $v$ and then accelerates until it comes back to the same speed in the opposite direction. 
The time interval in this process (in terms of the detector's proper time $\tau$) is given by $\Delta \tau=(2c/a)\text{arctanh}(v/c)$, and inserting $a=10^{20}$ m/s$^2$, it requires $v\approx c$ to yield $\Delta \tau \sim 1$ s. 
Otherwise, if we demand a slower speed, say $v=c/2$, then the interaction duration amounts to $\Delta \tau \sim 10^{-12}$ seconds, which is an extremely short period of time. 
The detector certainly cannot thermalize within such a time scale. 
Finally, executing such a protocol demands considerable space. 
The procedure in the Unruh QOE consists of linear acceleration and uniform motion at speed $v$. 
This means that it requires the detector to consistently move in one direction, which cannot be done in a confined laboratory.

Instead, one can consider a UDW detector in circular motion, a concept that is motivated by \cite{Bell.circular.1983, Bell.Leinaas.1987, BEC_Unruh}, where an experimental setup is proposed to measure the circular Unruh effect. 
See also \cite{Retzker.BEC.2008, Marino.zero-point.circular, CircularTemperaturesUnruh, bunney2023sound}. 
A circulating UDW detector overcomes the shortcomings described above, though the temperature induced by the acceleration is no longer $T\ts{U}$. 
In fact, the detector cannot be thermalized. 
Instead, one should define the effective temperature $T\ts{circ}$ as observed by the circulating detector \cite{CircularTemperaturesUnruh, Good.Unruh-like.effective.temp, Unruh.acceleration.rad.electron.1998.arxiv}. 
This is where our proposed framework excels, as our expressions remain valid even in the absence of thermality.

The outline of this paper is as follows. 
After we introduce the basic aspects of the UDW detector and thermality in quantum field theory in Sec.~\ref{sec:setup}, we review the mechanism of QOE in Sec.~\ref{subsec:the cycle}. 
Our main results are shown in Sec.~\ref{subsec:work extracted}, followed by examples in Sec.~\ref{subsec:inertial bath} and Sec.~\ref{subsec:circular}.

Throughout this paper, we use $\hbar = c= k\ts{B}=1$ and the mostly-positive signature convention: $(-+++)$.

\section{Setup}
\label{sec:setup}

\subsection{Unruh-DeWitt detector model}
\label{subsec:UDW}

Consider a pointlike UDW detector, which is a two-level quantum system with an energy gap $\Omega$ between ground $\ket{g}$ and excited states $\ket{e}$ locally interacting with the quantum Klein-Gordon field $\hat\phi(\sx)$. 
Here, the background spacetime is not restricted to a specific geometry but is assumed to be a (curved) spacetime that accommodates a well-behaved Klein-Gordon equation.

In the Schr\"odinger picture, the total Hamiltonian $\hat H\ts{S,tot}$ reads 
\begin{align}
    \hat H\ts{S,tot}
    &=
        \hat H\ts{S,D} + \hat H\ts{S,$\phi$} + \hat H\ts{S,int}\,,
\end{align}
where $\hat H\ts{S,D}$ and $\hat H\ts{S,$\phi$}$ are the free Hamiltonians for the detector and the field, respectively, and $\hat H\ts{S,int}$ is the interaction Hamiltonian. 
These are explicitly given by 
\begin{subequations}
    \begin{align}
        \hat H\ts{S,D}
        &=
            \Omega \ket{e}\bra{e} \otimes \id_\phi\,, \\
        \hat H\ts{S,$\phi$}
        &=
            \id\ts{D} \otimes  \int_{\R^n} \dd^n k\,\omega_{\bk} \hat a_{\bk}^\dag \hat a_{\bk}\,, \\
        \hat H\ts{S,int}
        &=
            \lambda \chi(\tau/\sigma) (\ket{e}\bra{g} + \ket{g}\bra{e}) \otimes \hat \phi(\bx(\tau))\,,
    \end{align}
\end{subequations}
where $\lambda$ is the coupling constant, $\tau$ is the proper time of the detector, and $\chi(\tau/\sigma)$ is the switching function (with $\sigma$ being the typical time scale of interaction), which determines the time dependence of coupling. 
In this paper, we assume that the $L^2$ norm of $\chi(\tau)$ is unity: 
\begin{align}
    ||\chi||_2
    \coloneqq
        \sqrt{ \int_\R \dd \tau\, |\chi(\tau)|^2 }
        =1\,.
\end{align}

Let $\hat H\ts{I}^\tau(\tau)$ be the interaction Hamiltonian in the interaction picture. 
Here, the Hamiltonian is the generator of time-translation with respect to the proper time $\tau$, which is indicated by the superscript. 
The explicit form is given by
\begin{align}
    \hat H\ts{I}^{ \tau } ( \tau )
    &=
        \lambda \chi(\tau/\sigma) \hat \mu(\tau) 
        \otimes \hat \phi(\sx(\tau))\,,
\end{align}
where $\hat \mu(\tau) $ is the monopole moment describing the internal dynamics of the detector: 
\begin{align}
    \hat \mu(\tau) 
    &=
        \ket{e} \bra{g} e^{ \ii \Omega \tau }
        +
        \ket{g} \bra{e} e^{ -\ii \Omega \tau }\,.
\end{align}

Let us obtain the final state of the detector. 
The time-evolution operator in the interaction picture is given by 
\begin{align}
    \hat U\ts{I}
    &=
        \mathcal{T}_\tau 
        \exp 
        \kako{
            -\ii \int_{\mathbb{R}} \dd \tau\,\hat H\ts{I}^\tau(\tau)
        } \,,
\end{align}
where $\mathcal{T}_\tau$ is a time-ordering symbol with respect to the proper time $\tau$. 
Carrying out the perturbative method, this operator can be expanded as 
\begin{subequations}
\begin{align}
    \hat U\ts{I}
    &=
        \id + \hat U\ts{I}^{(1)} + \hat U\ts{I}^{(2)} + \mathcal{O}(\lambda^3)\,,\\
    \hat U\ts{I}^{(1)}
    &=
        -\ii \int_{-\infty}^\infty \dd \tau\,\hat H\ts{I}^\tau(\tau)\,,\\
    \hat U\ts{I}^{(2)}
    &=
        - \int_{-\infty}^\infty \dd \tau
        \int_{-\infty}^{\tau} \dd \tau'\,
        \hat H\ts{I}^\tau(\tau) \hat H\ts{I}^\tau(\tau')\,.
\end{align}
\end{subequations}
Consider the initial state of the total system, 
\begin{align}
    \rho\ts{tot,0}
    &=
        \rho\ts{D,0}
        \otimes 
        \rho_\phi\,,
\end{align}
where $\rho_\phi$ is the initial state of the field and $\rho\ts{D,0}$ is the detector's initial state given by 
\begin{align}
    \rho\ts{D,0}
    &=
        p \ket{e}\bra{e} + (1-p) \ket{g} \bra{g}\,.
        \quad p\in [0,1] \label{eq:UDW initial}
\end{align}
The final total state $\rho\ts{tot}$ can be obtained as 
\begin{align}
    \rho\ts{tot}
    &=
        \hat U\ts{I} \rho\ts{tot,0} \hat U\ts{I}^\dag 
    =
        \sum_{m,n=0}^\infty \hat U\ts{I}^{(m)} \rho\ts{tot,0} \hat U\ts{I}^{(n)\dagger} 
\end{align}
and by tracing out the field degree of freedom we obtain
the final state of the detector, $\rho\ts{D}$.
In general, $\rho\ts{D}$ contains $n$-point correlation functions $\braket{\hat \phi(\sx_1)...\hat \phi(\sx_n)}_{\rho_\phi}$, where $\braket{ \cdot }_{\rho_\phi}$ is the expectation value with respect to $\rho_\phi$. 
In this paper, we are particularly interested in a quantum field in a quasi-free state,\footnote{Sometimes quasi-free states are called Gaussian states in the literature.} in which
the one-point correlator $\braket{\hat\phi(\sx)}_{\rho_\phi}$ vanishes and every correlation function can be written in terms of two-point correlators $W(\sx, \sx')\coloneqq \braket{\hat \phi(\sx) \hat \phi(\sx')}_{\rho_{\phi}}$, also known as the Wightman function. 
Examples of quasi-free states include the vacuum state $\ket{0}$ and the Kubo-Martin-Schwinger (KMS) state, which we will describe in the next subsection \ref{subsec:KMS}. 
Assuming that $\rho_\phi$ is a quasi-free state, the post-interaction state of the detector to the leading order in $\lambda$ can be calculated as follows. 
\begin{align}
    \rho\ts{D}
    &=
        \Tr_\phi[ \rho\ts{tot} ] 
    =
        (p + \delta p) \ket{e}\bra{e} + (1-p - \delta p) \ket{g}\bra{g} 
        +
        \mathcal{O}(\lambda^4)\,, \label{eq:final detector density}
\end{align}
where 
\begin{align}
    \delta p
    &=
        \lambda^2 \sigma
        \big[
            (1-p) \mathcal{F}(\Omega) 
            -
            p \mathcal{F}(-\Omega)
        \big]\,,\label{eq:delta p and response}
\end{align}
with 
\begin{align}
    &\mathcal{F}(\Omega) 
    =
        \dfrac{1}{\sigma}
        \int_\R \dd \tau 
        \int_\R \dd \tau'\,
        \chi(\tau/\sigma) \chi(\tau'/\sigma)
        e^{ -\ii \Omega (\tau - \tau') }
        W(\sx(\tau), \sx(\tau')) \label{eq:response function}
\end{align}
being the response function, {\it i.e.} the probability to excite the ground state $\ket{g}$ to $\ket{e}$. 
Similarly, $\mathcal{F}(-\Omega)$ is the probability for the transition $\ket{e} \to \ket{g}$. 
The quantity $W(\sx(\tau), \sx(\tau'))\coloneqq \braket{\hat \phi(\sx(\tau)) \hat \phi(\sx(\tau'))}_{\rho_{\phi}}$ is the pullback of the Wightman function along the detector's trajectory $\sx(\tau)$. 
Note that the odd-powers in $\lambda$ vanish since the one-point correlators $\braket{\hat \phi(\sx)}_{\rho_{\phi}}$ are zero.

We remark that $\rho\ts{D}$ is the state of the detector after interacting with the field once, while in QOE, the detector interacts with the field twice. 
Fortunately, Eq.~\eqref{eq:final detector density} tells us that the density matrix $\rho\ts{D}$ stays diagonal after the interaction, meaning the final density matrix also remains diagonal after the second interaction. 
Nevertheless, careful attention must be paid to the switching timings of these two interactions.

\begin{figure}[t]
\centering
\includegraphics[width=0.5\linewidth]{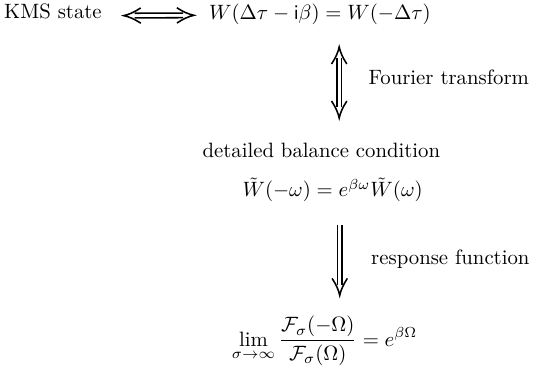}
\caption{The relationship between the KMS state and the response function.}
\label{fig:KMS fig}
\end{figure}

\subsection{KMS condition}\label{subsec:KMS}
In quantum mechanics, a thermal state is given by the Gibbs state $\rho=e^{-\beta \hat H}/Z$, where $\beta$ is the inverse temperature, $\hat H$ is the Hamiltonian, and $Z\coloneqq \Tr[e^{-\beta \hat H}]$ is the partition function. 
In quantum field theory, the notion of thermality is rigorously defined by the Kubo-Martin-Schwinger (KMS) condition \cite{Kubo1957thermality, Martin-Schwinger1959thermality, haag1967equilibrium}. 
The argument given in this section is summarized in Fig.~\ref{fig:KMS fig}.

Consider a Wightman function in a field state $\rho_\phi$: 
\begin{align}
    W(\tau, \tau')
    &=
        \braket{ \hat \phi(\sx(\tau)) \hat \phi(\sx(\tau')) }_{\rho_\phi}\,.
\end{align}
Note that the vacuum Wightman function $\braket{0|\hat \phi(\sx(\tau)) \hat \phi(\sx(\tau'))|0}$ is a special case corresponding to   $\rho_\phi=\ket{0}\bra{0}$. 
The Wightman function satisfies the KMS condition with respect to time $\tau$ at the KMS temperature $\beta\ts{KMS}^{-1} \equiv T\ts{KMS}$ if 
\begin{align}
    W(\tau - \ii \beta\ts{KMS}, \tau')
    &=
        W(\tau', \tau)\,.
\end{align}
In addition, if the KMS condition is satisfied, then the Wightman function is stationary, i.e., it only depends on the difference in time: $W(\tau, \tau')=W(\Delta \tau)$. 
We identify the KMS temperature $T\ts{KMS}$ as the temperature of the quantum field.

The KMS condition is related to the so-called detailed balance condition. 
For any function $f(\omega)$, the detailed balance condition at temperature $\beta^{-1}$ is given by 
\begin{align}
    f(-\omega)= e^{\beta \omega} f(\omega)\,. \label{eq:detailed balance cond}
\end{align}
It is known \cite{Fewster.Waiting.Unruh} that, assuming the Wightman function is stationary, the KMS condition with respect to $\tau$ at temperature $\beta\ts{KMS}^{-1}$ is satisfied if and only if the Fourier transformed Wightman function, 
\begin{align}
    \tilde{W}(\omega)
    &\coloneqq
        \int_\R \dd \Delta \tau\,
        W(\Delta \tau) e^{ -\ii \omega \Delta \tau }\,,
\end{align}
satisfies the detailed balance condition \eqref{eq:detailed balance cond}.


The detailed balance condition can be implemented into the response function of a UDW detector. 
Suppose the Wightman function satisfies the KMS condition and the switching function $\chi(\tau/\sigma)$ has a typical interaction duration time scale $\sigma$. 
The response function \eqref{eq:response function} can be written as 
\begin{align}
    \mathcal{F}_\sigma (\Omega)
    &=
        \int_\R \dfrac{ \dd \bar \omega }{ 2\pi } \,
        |\tilde \chi(\bar\omega)|^2 
        \tilde W(\Omega + \bar \omega/ \sigma)\,, \label{eq:response function Fourier}
\end{align}
where $\bar \omega$ is a dimensionless variable and $\tilde{\chi}(\bar\omega)$ is the Fourier transform of the switching function $\chi(\tau/\sigma)$. 
One can show that if $\tilde{\chi}(\bar\omega)$ decays sufficiently fast as $\bar\omega$ increases (such as a Gaussian switching), then the response function satisfies the detailed balance condition in the long interaction limit \cite{Fewster.Waiting.Unruh, Garay2016anti-unruh}: 
\begin{align}
    \lim_{\sigma \to \infty} 
    \dfrac{\mathcal{F}_\sigma(-\Omega)}{ \mathcal{F}_\sigma(\Omega) } 
    = 
        e^{\beta\ts{KMS} \Omega}\,. \label{eq:detailed balance for response}
\end{align}

Inspired by the detailed balance condition for the response function \eqref{eq:detailed balance for response}, one may define the \textit{effective temperature} $T\ts{eff}$: 
\begin{align}
    T\ts{eff}^{-1}
    &\coloneqq
        \dfrac{1}{\Omega} 
        \ln \dfrac{\mathcal{F}_\sigma(-\Omega)}{ \mathcal{F}_\sigma(\Omega) } \,. \label{eq:effective temperature}
\end{align}
The effective temperature is an estimator for the KMS temperature of the field. 
In particular, if the Wightman function obeys the KMS condition and the long interaction limit ($\sigma \to \infty$) is taken, then the effective temperature becomes the KMS temperature due to \eqref{eq:detailed balance for response}.

\begin{figure*}[tp]
\centering
\includegraphics[width=\textwidth]{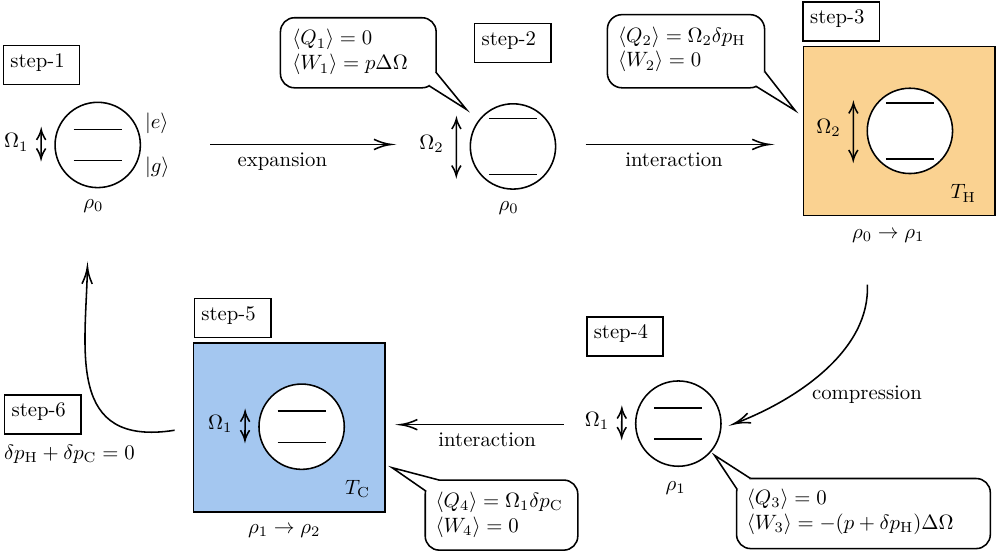}
\caption{The quantum Otto cycle. }
\label{fig:Otto cycle}
\end{figure*}

\section{Relativistic quantum Otto engine}\label{sec: QOE}

\subsection{Review: quantum Otto engine}\label{subsec:the cycle}

In this subsection, we review the QOE following the paper by Kieu \cite{Kieu.secondlaw.demon.otto}. 
Note that the whole system is composed of a qubit and thermal baths in the Gibbs state. 
In the next subsection, we will take the following protocol and apply it to the system composed of a UDW detector interacting with a quantum scalar field in a quasi-free state, namely, we will omit the assumption that the baths are in a thermal state.

\begin{enumerate}
 \item Prepare a qubit with energy gap $\Omega_1$ in an initial state $\rho_0= p \ket{e}\bra{e} + (1-p) \ket{g} \bra{g}$, where $p \in [0,1]$, as shown in Fig.~\ref{fig:Otto cycle}. 
 \item Let the qubit undergo an adiabatic expansion (zero heat exchange: $\braket{Q_1}=0$) to enlarge the energy gap from $\Omega_1$ to $\Omega_2$, where $\Omega_2 > \Omega_1$. 
 The Hamiltonian of the qubit is given by $\hat H(t)=\Omega(t) \ket{e}\bra{e}$ and so the work done on the qubit is 
 \begin{align}
     \braket{W_1}
     &=
        \int \dd t\,
        \Tr \kagikako{
            \rho_0 \dfrac{ \dd \hat H(t) }{\dd t}
        }
    =
        p \Delta \Omega\,,
 \end{align}
 where $\Delta \Omega \coloneqq \Omega_2 - \Omega_1$. 
 The qubit's state remains as  $\rho_0$ due to the quantum adiabatic theorem. 
 \item Let the qubit interact with a bath at temperature $T\ts{H}$. 
 The qubit's Hamiltonian is time-independent $\hat H=\Omega_2 \ket{e}\bra{e}$, and so the work done on the qubit is zero: $\braket{W_2}=0$. 
 After the interaction, the state becomes $\rho_1= (p + \delta p\ts{H}) \ket{e}\bra{e} + (1- p -\delta p\ts{H}) \ket{g} \bra{g}$. 
 The first law of thermodynamics gives the dissipated heat: 
 \begin{align}
     \braket{Q_2}
     &=
        \Tr[\rho_1 \hat H] - \Tr[\rho_0 \hat H]
    =
        \Omega_2 \delta p\ts{H}\,.
 \end{align}
 This is an isochoric thermalization process. 
 \item After isolating the qubit from the bath, perform an adiabatic contraction of the energy gap from $\Omega_2 $ to $\Omega_1$, while the state remains to be $\rho_1$. 
 From the same calculation as in step-2, we have 
 \begin{align}
     \braket{Q_3}
     &=
        0\,, \\
    \braket{W_3}
    &=
        -( p + \delta p\ts{H}) \Delta \Omega\,.
 \end{align}
 The work is extracted from the qubit in this process. 
 \item Next another isochoric thermalization process is carried out:  the qubit  interacts with a bath at a colder temperature $T\ts{C} (< T\ts{H})$. 
 At the end of the process, the state becomes $\rho_2=(p + \delta p\ts{H} + \delta p\ts{C}) \ket{e}\bra{e} + (1 - p - \delta p\ts{H} - \delta p\ts{C}) \ket{g}\bra{g}$ and so 
 \begin{align}
     \braket{Q_4}
     &=
        \Omega_1 \delta p\ts{C}\,,\\
    \braket{W_4}
    &=
        0\,.
 \end{align}
 The total extracted work and dissipated heat read
 \begin{align}
     \braket{W\ts{ext}}
     &=
        -\sum_{j=1}^4 \braket{W_j}
    =
        \delta p\ts{H} \Delta \Omega\,, \label{eq:total extracted work} \\
    \braket{Q}
    &=
        \sum_{j=1}^4 \braket{Q_j}
    =
        \Omega_2 \delta p\ts{H} + \Omega_1 \delta p\ts{C}\,. \label{eq:total heat}
 \end{align}
 Assuming $\Omega_2 > \Omega_1$, the extracted work is positive if $\delta p\ts{H}>0$. 
 \item Impose $\delta p\ts{H} + \delta p\ts{C}=0$ to complete a thermodynamic cycle. 
 From this condition, Eqs.~\eqref{eq:total extracted work} and \eqref{eq:total heat} yield $\braket{Q} = \braket{W\ts{ext}}$, which obeys the first law of thermodynamics. 
\end{enumerate}

It is known that the positive work condition reads \cite{Feldmann.QOE.2000, Kieu.secondlaw.demon.otto} $T\ts{C}/\Omega_1 < T\ts{H}/\Omega_2$ and the efficiency of the QOE is $\eta\ts{O} \coloneqq \braket{W\ts{ext}}/Q_2 =1-\Omega_1/\Omega_2$.

\begin{figure}[t]
\centering
\includegraphics[width=0.5\linewidth]{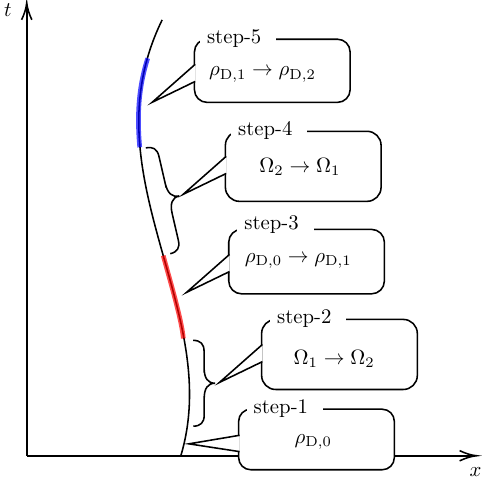}
\caption{Each step in a relativistic QOE. 
The red and blue stripes indicate the isochoric process, i.e., the interaction between the detector and the field.}
\label{fig:relativistic QOE}
\end{figure}

\subsection{Work extracted by a UDW detector}
\label{subsec:work extracted}
Now let us employ a UDW detector as the qubit in this process. 
As mentioned before, we will not assume that the baths are thermal. 
Instead, we only assume that the quantum field is in a quasi-free state, i.e., the one-point correlators $\braket{\hat \phi(\sx)}_{\rho_{\phi}}$ vanish. 
Nevertheless, we will keep using subscripts such as $\delta p\ts{H}$ and $\delta p\ts{C}$ for convenience.

In this subsection, we consider a general scenario in which a UDW detector following an arbitrary path in spacetime interacts with a quantum field in a quasi-free state. 
Working perturbatively, we derive the extracted work and the condition for it to be positive.

\subsubsection{Relativistic quantum Otto engine}

We first describe our QOE in the relativistic setting as depicted in Fig.~\ref{fig:relativistic QOE}. 
\begin{itemize}
 \item \textit{(step-1)} Before the interaction, the detector with the energy gap $\Omega_1$ is prepared in $\rho\ts{D,0}$ in \eqref{eq:UDW initial}. 
 \item \textit{(Step-2)} One performs a quantum adiabatic expansion $\Omega_1\to \Omega_2$ with $\Omega_2> \Omega_1$. 
During this process, the detector does not interact with the field. 
 \item \textit{(Step-3)} Isochoric process. 
The detector interacts with the field with a time-dependence specified by a compactly supported switching function $\chi\ts{H}[(\tau-\tau\ts{H})/\sigma\ts{H}]$, where $\tau\ts{H}$ is the center of interaction and $\sigma\ts{H}(>0)$ is the interaction duration; its  support  is $\text{supp}(\chi\ts{H}) = [\tau\ts{H} - \sigma\ts{H}/2, \tau\ts{H} + \sigma\ts{H}/2]$. 
The state changes from $\rho\ts{D,0}\to \rho\ts{D,1}$, where $\rho\ts{D,1}$ is given by Eqs.~\eqref{eq:final detector density}-\eqref{eq:response function}, that is, 
\begin{subequations}
    \begin{align}
        &\rho\ts{D,1}
        =
            (p + \delta p\ts{H}) \ket{e}\bra{e} + (1-p - \delta p\ts{H}) \ket{g}\bra{g} 
            +
            \mathcal{O}(\lambda^4)\,, \\
        &\delta p\ts{H}
        =
            \lambda^2 \sigma\ts{H}
            \big[
                (1-p) \mathcal{F}\ts{H}(\Omega_2) 
                -
                p \mathcal{F}\ts{H}(-\Omega_2)
            \big]\,, \\
        &\mathcal{F}\ts{H}(\Omega_2)
        =
            \dfrac{1}{\sigma\ts{H}}
            \int_\R \dd \tau 
            \int_\R \dd \tau'\,
            \chi\ts{H}\left( \frac{ \tau-\tau\ts{H} }{ \sigma\ts{H} } \right) 
            \chi\ts{H}\left( \frac{ \tau'-\tau\ts{H} }{ \sigma\ts{H} } \right) 
            e^{ -\ii \Omega_2 (\tau - \tau') }
            W(\sx(\tau), \sx(\tau'))\,.
    \end{align}
\end{subequations}
While we identified the switching function as a compactly supported function, it is also possible to employ rapidly decreasing functions such as   Gaussian or Lorentzian functions. 
These are not compactly supported, but the results remain unchanged when the exponentially decreasing tails of these functions are disregarded. 
We will make use of these functions later in this subsection. 
 \item \textit{(Step-4)} Adiabatic compression $\Omega_2 \to \Omega_1$. 
Again, the detector does not interact during this process and so the state remains as $\rho\ts{D,1}$. 
 \item \textit{(Step-5)} Implement another isochoric process using $\chi\ts{C}[(\tau - \tau\ts{C})/\sigma\ts{C}]$ for the switching function.
To ensure that the interaction does not overlap with the previous isochoric process, we make sure that the supports of the two switching functions are disjoint by implementing $\tau\ts{H}+\sigma\ts{H}/2 < \tau\ts{C}-\sigma\ts{C}/2$. 
The state of the detector changes as $\rho\ts{D,1} \to \rho\ts{D,2}$ where $\rho\ts{D,2}=(p + \delta p\ts{H} + \delta p\ts{C}) \ket{e}\bra{e} + (1 - p - \delta p\ts{H} - \delta p\ts{C}) \ket{g}\bra{g}$. 
The quantity $\delta p\ts{C}$ is 
\begin{subequations}
    \begin{align}
        &\delta p\ts{C}
        =
            \lambda^2 \sigma\ts{C}
            \big[
                (1-p - \delta p\ts{H}) \mathcal{F}\ts{C}(\Omega_1) 
                - 
                (p + \delta p\ts{H}) \mathcal{F}\ts{C}(-\Omega_1)
            \big] 
            + \mathcal{O}(\lambda^6) \,, \\
        &\mathcal{F}\ts{C}(\Omega_1)
            =
                \dfrac{1}{\sigma\ts{C}}
                \int_\R \dd \tau 
                \int_\R \dd \tau'\,
                \chi\ts{C}\left( \frac{ \tau-\tau\ts{C} }{ \sigma\ts{C} } \right) 
                \chi\ts{C}\left( \frac{ \tau'-\tau\ts{C} }{ \sigma\ts{C} } \right) 
                e^{ -\ii \Omega_1 (\tau - \tau') }
                W(\sx(\tau), \sx(\tau'))\,.
    \end{align}
\end{subequations}
Notice that $\delta p\ts{C}$ contains $\lambda^4$ terms since $\delta p\ts{H}$ is of order $\lambda^2$. 
As in the traditional QOE, the work extracted reads $\braket{W\ts{ext}}=\delta p\ts{H} \Delta \Omega$. 
 \item \textit{(Step-6)} Finally, we impose the condition $\delta p\ts{H}+\delta p\ts{C}=0$ to complete a thermodynamic cycle. 
\end{itemize}

Since $\braket{W\ts{ext}}$ only depends on $\delta p\ts{H}$, it may appear that the cyclicity condition does not affect this value. 
However there is a subtlety. 
If we assume that an experimenter can freely choose a value of $p\in [0,1]$ at the beginning of the cycle, this means that the experimenter has to adjust the response functions $\mathcal{F}\ts{H}(\pm \Omega_2)$ and $\mathcal{F}\ts{C}(\pm \Omega_1)$ in such a way that the cyclicity condition is satisfied. 
On the other hand, if we allow the response functions to take any value, then the population $p$ has to be adjusted accordingly.

\subsubsection{When $p$ is free to choose}\label{subsubsec:when p is free}

We first require the cyclicity condition, $\delta p\ts{H} + \delta p\ts{C}=0$, to close the thermodynamic cycle. 
Since $\delta p\ts{C}$ contains terms of order $\mathcal{O}(\lambda^4)$, we omit these by imposing $1 \gg \delta p\ts{H}/p$ for a given $p(\neq 0)$. 
This reduces to 
\begin{align}
    p 
    \gg 
        \dfrac{ \lambda^2 \sigma\ts{H} \mathcal{F}\ts{H}(\Omega_2) }
        { 1 + \lambda^2 \sigma\ts{H} [ \mathcal{F}\ts{H}(\Omega_2) + \mathcal{F}\ts{H}(-\Omega_2) ] }\,.
\end{align}
In particular, if the long-interaction limit $\sigma\ts{H}\to \infty$ is taken, we obtain 
\begin{align}
    p 
    \gg 
        \dfrac{ 1 }
        { 1 + \displaystyle\lim_{\sigma\ts{H} \to \infty } \mathcal{F}\ts{H}(-\Omega_2)/\mathcal{F}\ts{H}(\Omega_2) }
    =
        \dfrac{ 1 }
        { 1 + \displaystyle\lim_{\sigma\ts{H} \to \infty } e^{ \Omega_2 / T\ts{H}^{\text{eff}} } }\,.
\end{align}
Thus, in the long interaction duration, $T\ts{H}^{\text{eff}}/\Omega_2 \ll 1$ is required. 
Moreover, as a special case of a detector with a rapidly decreasing switching function interacting with the quantum field in the KMS state, this reads 
\begin{align}
    p 
    \gg 
        \dfrac{ 1 }
        { 1 + e^{ \Omega_2 / T\ts{H}^{\text{KMS}} } }\,.
\end{align}
Nevertheless, if the interaction duration is finite, then $\lambda \ll 1$ is sufficient for our perturbative analysis.\footnote{To be precise, one has to make $\lambda$ a dimensionless quantity since it has units of $[\text{Length}]^{(n-3)/2}$ in $(n+1)$ dimensions.}

Under this restriction, the cyclicity condition for any $p\in (0,1]$ leads to 
\begin{align}
    \sigma\ts{C} \mathcal{F}\ts{C}(\Omega_1) 
    \kako{
        \dfrac{1-p}{p}
        - 
        \dfrac{ \mathcal{F}\ts{C}(-\Omega_1) }{ \mathcal{F}\ts{C}(\Omega_1) }
    }
    =
    - 
    \sigma\ts{H} \mathcal{F}\ts{H}(\Omega_2) 
    \kako{
        \dfrac{1-p}{p}
        - 
        \dfrac{ \mathcal{F}\ts{H}(-\Omega_2) }{ \mathcal{F}\ts{H}(\Omega_2) }
    }\,. \label{eq:free p cyclicity condition}
\end{align}
That is, one has to tune the response functions so that this equality is satisfied.

We then consider the positive work condition in the case where the value of $p$ is chosen freely by an experimenter. 
The work extracted \eqref{eq:total extracted work} is 
\begin{align}
    \braket{W\ts{ext}}
    &=
        \lambda^2 \sigma\ts{H}
        \big[
            (1-p) \mathcal{F}\ts{H}(\Omega_2) 
            -
            p \mathcal{F}\ts{H}(-\Omega_2) 
        \big]
        \Delta \Omega\,, \label{eq:free p work extracted}
\end{align}
and the positive work condition is $(1-p) \mathcal{F}\ts{H}(\Omega_2) - p \mathcal{F}\ts{H}(-\Omega_2) >0$, i.e., $(1-p)/p > \mathcal{F}\ts{H}(-\Omega_2)/\mathcal{F}\ts{H}(\Omega_2)$. 
We note that $\mathcal{F}\ts{H}(-\Omega_2)/\mathcal{F}\ts{H}(\Omega_2) > 1$ since the de-excitation probability is higher than the excitation probability, thereby implying $p \in (0,1/2)$. 
This is consistent with   previous work \cite{UnruhOttoEngine}, where the Unruh QOE with a free $p$ is able to extract work only when $0<p<1/2$. 
The condition can be written in terms of the effective temperature \eqref{eq:effective temperature}: 
\begin{align}
    \kako{
        \ln \dfrac{ 1-p }{p}
    }^{-1} 
    < 
    \dfrac{T\ts{H}^{\text{eff}}}{ \Omega_2 }\,,
    \quad (0< p < 1/2)\,. \label{eq:free p positive work condition}
\end{align}

It should be recalled that in the long interaction duration, $T\ts{H}^{\text{eff}}/\Omega_2 \ll 1$ is required, which means that \eqref{eq:free p positive work condition} tells us that $p \ll 1$ needs to be chosen if we wish to extract positive work in the long-interaction limit within the perturbation theory.

Notice that if positive work is extracted while the cyclicity condition is satisfied, then $\mathcal{F}\ts{C}(\pm \Omega_1)$ has to obey $(1-p)/p < \mathcal{F}\ts{C}(-\Omega_1)/\mathcal{F}\ts{C}(\Omega_1)$ since the response functions are non-negative and the positive work condition requires $(1-p)/p > \mathcal{F}\ts{H}(-\Omega_2)/\mathcal{F}\ts{H}(\Omega_2)$. 
Therefore, within the perturbation theory, 
\begin{align}
    \braket{W\ts{ext}}>0~\text{from QOE} 
    \Rightarrow 
        \dfrac{T\ts{C}^{\text{eff}}}{\Omega_1}
        <
        \kako{
            \ln \dfrac{ 1-p }{p}
        }^{-1} 
        <
        \dfrac{T\ts{H}^{\text{eff}}}{\Omega_2}  \,. \label{eq:free p necessary condition}
\end{align}
The opposite is not always true since the inequality does not necessarily imply the cyclicity condition \eqref{eq:free p cyclicity condition}. 
In addition, the statement \eqref{eq:free p necessary condition} does not hold without the cyclicity condition. 
Nevertheless, by contraposition, if $T\ts{H}^{\text{eff}}/\Omega_2 > T\ts{C}^{\text{eff}}/\Omega_1$ is violated then positive work cannot be extracted \textit{or} the cycle is not closed.

The inequality $T\ts{H}^{\text{eff}}/\Omega_2 > T\ts{C}^{\text{eff}}/\Omega_1$ generalizes the positive work condition found in \cite{Feldmann.QOE.2000, Kieu.secondlaw.demon.otto}, by replacing bath temperatures with effective temperatures. 
Specifically, effective temperatures reduce to the KMS temperatures when the pullback of the Wightman function along a trajectory satisfies the KMS condition and when the detector, with a rapidly decreasing switching function, interacts with the field for a long time. 
In this particular scenario, the work extracted becomes 
\begin{align}
    \braket{W\ts{ext}}
    &\approx
        \lambda^2 \sigma\ts{H} \mathcal{F}\ts{H}(\Omega_2) 
        \kagikako{
            1
            -
            p 
            \kako{
                1 + e^{\Omega_2/T\ts{H}^{\text{KMS}}}
            }
        }
        \Delta \Omega\,. \label{eq:free p asymptotic work extracted}
\end{align}

\subsubsection{When $p$ is adjusted}\label{subsubsec:when p is adjusted}

Let us focus on the case where $p$ is initially adjusted and derive the work extracted. 
We again assume that $p\neq 0$ and $p\gg \delta p\ts{H}$ so that the terms higher than $\lambda^4$ can be omitted in $\delta p\ts{C}$. 
From the cyclicity condition, we can solve for $p$, which reads 
\begin{align}
    p
    &=
        \dfrac{ \sigma\ts{H} \mathcal{F}\ts{H}(\Omega_2) + \sigma\ts{C} \mathcal{F}\ts{C}(\Omega_1) }
        { \sigma\ts{H} [\mathcal{F}\ts{H}(\Omega_2) + \mathcal{F}\ts{H}(-\Omega_2) ] + \sigma\ts{C} [\mathcal{F}\ts{C}(\Omega_1) + \mathcal{F}\ts{C}(-\Omega_1) ] }\,.\label{eq:p after cyclic condition}
\end{align}
By inserting $p$ into \eqref{eq:total extracted work} we obtain 
\begin{align}
    &\braket{W\ts{ext}}
    = 
        \lambda^2 \sigma\ts{H} \sigma\ts{C}
        \dfrac{ \mathcal{F}\ts{H}(\Omega_2) \mathcal{F}\ts{C}(-\Omega_1) - \mathcal{F}\ts{H}(-\Omega_2) \mathcal{F}\ts{C}(\Omega_1) }
        { \sigma\ts{H} [ \mathcal{F}\ts{H}(\Omega_2) + \mathcal{F}\ts{H}(-\Omega_2) ] + \sigma\ts{C} [ \mathcal{F}\ts{C}(\Omega_1) + \mathcal{F}\ts{C}(-\Omega_1) ] }
        \Delta \Omega\,. \label{eq:extracted work wrt response}
\end{align}
The extract work is positive if and only if
\begin{align}
    \mathcal{F}\ts{H}(\Omega_2) \mathcal{F}\ts{C}(-\Omega_1) > \mathcal{F}\ts{H}(-\Omega_2) \mathcal{F}\ts{C}(\Omega_1)\,.\label{eq:positive work condition wrt response}
\end{align}
We find that the positive work condition \eqref{eq:positive work condition wrt response} can be written in terms of the effective temperature defined by \eqref{eq:effective temperature} as 
\begin{align}
    \dfrac{T\ts{C}^{\text{eff}} }{ \Omega_1 }
    < 
    \dfrac{T\ts{H}^{\text{eff}} }{ \Omega_2 }\,.\label{eq:positive work condition wrt effective temp}
\end{align}
Again, we are not necessarily assuming that the field is in the KMS state, and so the effective temperature is not necessarily the KMS temperature. 
If the field is in the KMS state and the detector with a rapidly decreasing switching function interacts with the field for a long time ($\sigma\ts{H}, \sigma\ts{C} \to \infty$), the effective temperature becomes the KMS temperature, leading to 
\begin{align}
    \dfrac{T\ts{C}^{\text{KMS}} }{ \Omega_1 } < \dfrac{T\ts{H}^{\text{KMS}} }{ \Omega_2 }\,,\label{eq:positive work condition wrt temp}
\end{align}
which is the condition found in \cite{Feldmann.QOE.2000, Kieu.secondlaw.demon.otto}. 
The extracted work in this case is 
\begin{align}
    &\braket{W\ts{ext}}
    = 
        \lambda^2 \sigma\ts{H} \sigma\ts{C}
        \dfrac{ (e^{\Omega_1/T\ts{C}} - e^{\Omega_2/T\ts{H}} ) \mathcal{F}\ts{H}(\Omega_2) \mathcal{F}\ts{C}(\Omega_1) }
        { \sigma\ts{H}(1+e^{\Omega_2/T\ts{H}})\mathcal{F}\ts{H}(\Omega_2) + \sigma\ts{C} (1+e^{\Omega_1/T\ts{C}})\mathcal{F}\ts{C}(\Omega_1) }
        \Delta \Omega\,.\label{eq:work in detailed balanced}
\end{align}
Therefore, \eqref{eq:positive work condition wrt temp} is a special case of \eqref{eq:positive work condition wrt response} [and equivalently \eqref{eq:positive work condition wrt effective temp}] when the response function obeys the detailed balance condition \eqref{eq:detailed balance for response}. 
In other words, the condition \eqref{eq:positive work condition wrt response} (or alternatively  \eqref{eq:positive work condition wrt effective temp}) is applicable to any scenario, even when the detector is not thermalized.

In summary, we showed that within the perturbation theory, the positive work condition is associated with $T\ts{C}^{\text{eff}}/\Omega_1 < T\ts{H}^{\text{eff}}/\Omega_2$. 
If $p$ is not fixed and the cycle is closed by adjusting the response functions, $T\ts{C}^{\text{eff}}/\Omega_1 < T\ts{H}^{\text{eff}}/\Omega_2$ is a necessary condition to extract positive work. 
On the other hand, if we instead adjust $p$ to close the cycle while the response functions are not fixed, then this condition is necessary and sufficient. 
This recovers the traditional QOE positive work condition \cite{Feldmann.QOE.2000, Kieu.secondlaw.demon.otto} as a special case. 
Note that our results hold not only for a quantum field in the KMS state but also for any quasi-free state in curved spacetime. 
The use of the effective temperature is similar to other papers such as \cite{Huang.squeezing.work}.

\section{Examples}
We now demonstrate our results in $(3+1)$-dimensional Minkowski spacetime. 
Throughout this section, we use a Gaussian switching function 
\begin{align}
    \chi[(\tau-\tau_j)/\sigma)]
    &=
        e^{ -(\tau-\tau_j)^2/2\sigma^2 }\,,
    \quad j\in \{ \text{H}, \text{C} \} \label{eq:Gaussian switch}
\end{align}
where $\tau_j$ is the center of the Gaussian and $\sigma > 0$ is the characteristic Gaussian width, which has the units of time.

\subsection{Inertial detector in thermal quantum field}\label{subsec:inertial bath}
Consider $(3+1)$-dimensional Minkowski spacetime. 
We wish to examine the extracted work \eqref{eq:extracted work wrt response} by coupling an inertial UDW detector to a KMS state of the field to explore the basic properties of the work extraction.

The Wightman function in a thermal state of a massless field can be written as \cite{simidzija2018harvesting}
\begin{align}
    W_\beta(\sx, \sx')
    &=
        W\ts{M}(\sx, \sx') + W\ts{th}(\sx, \sx')\,,
\end{align}
where $W\ts{M}$ is the Wightman function in the Minkowski vacuum and $W\ts{th}$ is the contribution from the thermal state: 
\begin{subequations}
    \begin{align}
        W\ts{M}(\sx, \sx')
        &=
            -\dfrac{1}{4\pi^2} 
            \dfrac{1}{ (t-t'-\ii \epsilon)^2 - |\bx - \bx'|^2 }\,, \label{eq:Wightman Minkowski} \\
        W\ts{th}(\sx, \sx')
        &=
            \int_{\R^3} \dfrac{\dd^3 k}{ (2\pi)^3 2\kk }
            \dfrac{ e^{\ii \kk (t-t')-\ii \bk\cdot (\bx - \bx') } + \text{c.c.} }{ e^{\beta \kk} -1 }\,,
    \end{align}
\end{subequations}
Here, $\epsilon$ is the UV cutoff and $\beta=T^{-1}$ is the inverse KMS temperature of the field.

Assuming the Gaussian switching \eqref{eq:Gaussian switch}, the response function $\mathcal{F}_\beta(\Omega)$ of an inertial detector in the thermal bath reads \cite{simidzija2018harvesting}
\begin{align}
    \mathcal{F}_\beta(\Omega)
    &=
        \mathcal{F}\ts{M}(\Omega) 
        +
        \mathcal{F}\ts{th} (\Omega)\,,
\end{align}
where 
\begin{subequations}
    \begin{align}
        \mathcal{F}\ts{M}(\Omega)
        &=
            \dfrac{1}{4 \pi \sigma}
            \kagikako{
                e^{ -\Omega^2 \sigma^2 }
                - 
                \sqrt{\pi} \Omega \sigma \erfc(\Omega \sigma)
            }\,, \\
        \mathcal{F}\ts{th}(\Omega)
        &=
            \dfrac{ e^{-\Omega^2 \sigma^2} \sigma }{\pi}
            \int_0^\infty \dd k\,
            \dfrac{ k e^{-k^2 \sigma^2} \cosh (2\Omega \sigma^2 k) }{ e^{\beta k} - 1 }
    \end{align}
\end{subequations}
are respectively the response functions for an inertial detector in the Minkowski vacuum and the thermal contribution. 
Note that it does not depend on the center of the Gaussian switching since the Wightman function is time-translation invariant. 
Below, we assume that the interaction durations are the same: $\sigma\ts{H}=\sigma\ts{C} \equiv \sigma$.

\begin{figure}[t]
\centering
\includegraphics[width=\linewidth]{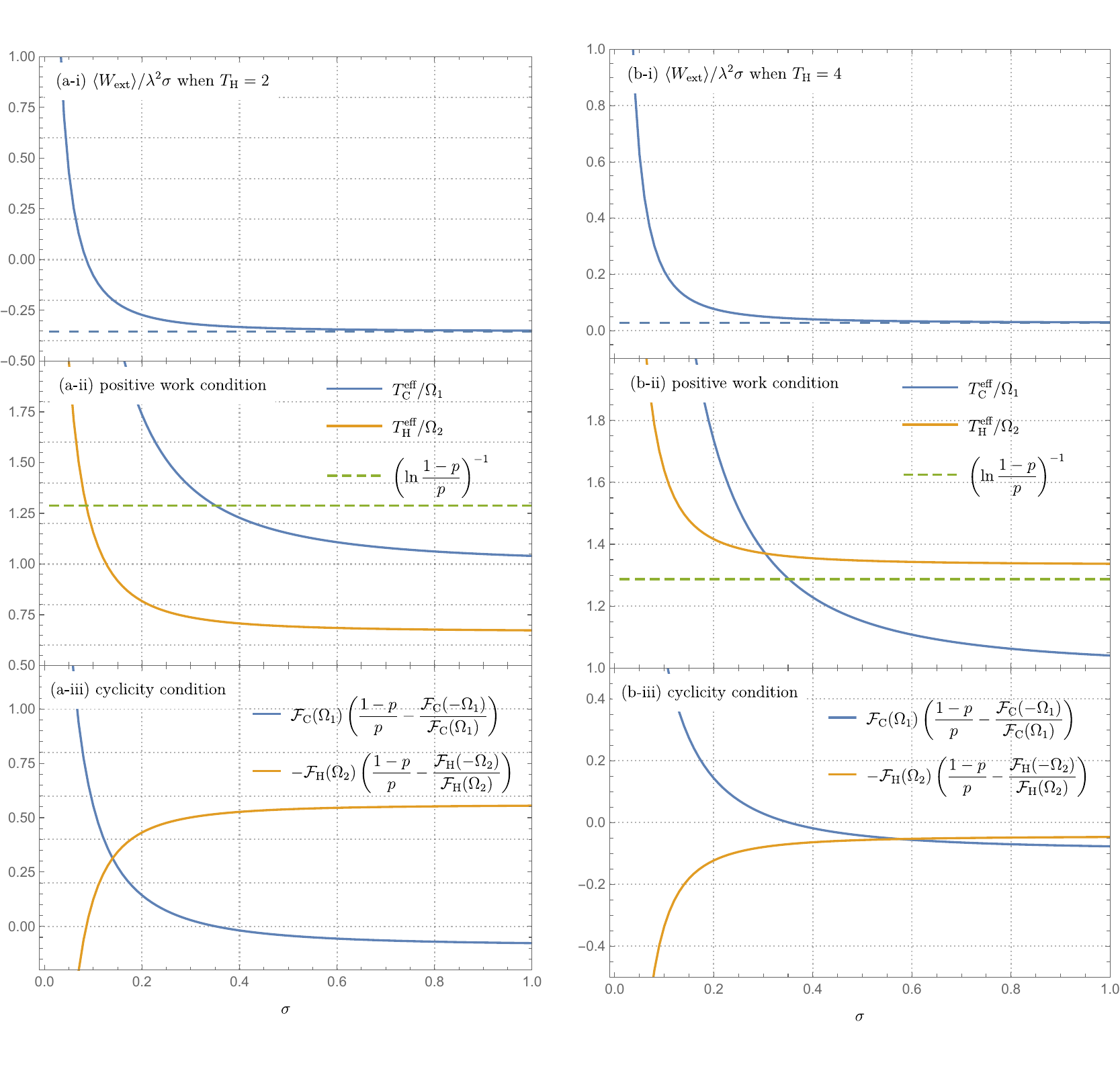}
\caption{The extracted work and conditions for a UDW detector in a KMS state of a thermal quantum field for (a) $T\ts{H}=2$ and (b) $T\ts{H}=4$. 
Here, $\Omega_2=3, T\ts{C}=1$, and $p=0.315$ in units of $\Omega_1$. 
(a-i) and (b-i) depict $\braket{W\ts{ext}}/\lambda^2 \sigma$ as a function of the interaction duration $\sigma$ in units of $\Omega_1$. 
The dashed line indicates the asymptotic value. 
The positive work condition is examined in (a-ii) and (b-ii), and the cyclicity condition is investigated in (a-iii) and (b-iii).  Negative values mean that work is done \textit{on} the system (i.e., the UDW detector).
}
\label{fig:FigThermalFreeP}
\end{figure}

\subsubsection{Case: free choice of $p$}

We first consider the case where $p$ is freely chosen as discussed in Sec.~\ref{subsubsec:when p is free}. 
We plot $\braket{W\ts{ext}}/\lambda^2 \sigma$ defined in \eqref{eq:free p work extracted} as a function of interaction duration $\sigma$ when $T\ts{H}=2, \Omega_2=3, p=0.315$, and $\Omega_1=1$ in Fig.~\ref{fig:FigThermalFreeP}(a-i). 
The dashed line represents the asymptotic value, $\lim_{\sigma \to \infty} \braket{W\ts{ext}}/\lambda^2 \sigma$, given by \eqref{eq:free p asymptotic work extracted}. 
We see that the work can be extracted when the interaction duration is short, but as $\sigma$ increases extraction becomes exponentially difficult. 
This behavior can be explained by examining the positive work condition \eqref{eq:free p positive work condition} in Fig.~\ref{fig:FigThermalFreeP}(a-ii). 
The figure shows that the positive work condition is satisfied when $\sigma$ is small enough. 
When $\sigma \to \infty$ is taken, the effective temperature becomes the KMS temperature: $T\ts{H}^{\text{eff}}/\Omega_2 \to T\ts{H}/\Omega_2=2/3$ in \ref{fig:FigThermalFreeP}(a-ii). 
Moreover, based on the discussion in Sec.~\ref{subsubsec:when p is free}, the perturbative analysis is valid in this limit when $T\ts{H}/\Omega_2 \ll 1$, which is not the case in our scenario. 
Nevertheless, the perturbative analysis is valid in the figure as long as $\lambda \ll 1$.

Although $\braket{W\ts{ext}}/\lambda^2 \sigma > 0$ for some $\sigma$, the thermodynamic cycle is not closed. 
The cyclicity condition \eqref{eq:free p cyclicity condition} when $\sigma\ts{H}=\sigma\ts{C}$ reads 
\begin{align}
    \mathcal{F}\ts{C}(\Omega_1) 
    \kako{
        \dfrac{1-p}{p}
        - 
        \dfrac{ \mathcal{F}\ts{C}(-\Omega_1) }{ \mathcal{F}\ts{C}(\Omega_1) }
    }
    =
    - 
    \mathcal{F}\ts{H}(\Omega_2) 
    \kako{
        \dfrac{1-p}{p}
        - 
        \dfrac{ \mathcal{F}\ts{H}(-\Omega_2) }{ \mathcal{F}\ts{H}(\Omega_2) }
    }\,. 
\end{align}
We plot each side of this equation in Fig.~\ref{fig:FigThermalFreeP}(a-iii),
 fixing $T\ts{C}=1$ and other parameters (except $\sigma$), so as to find an optimal interaction duration $\sigma$ such that the two curves cross. 
This point in the figure represents the scenario when the cyclicity condition is satisfied. 
As one can see, the positive work condition in \ref{fig:FigThermalFreeP}(a-ii) and the cyclicity condition in \ref{fig:FigThermalFreeP}(a-iii) are not satisfied at the same time; hence the work extracted is not coming from a genuine QOE. 
In fact, Fig.~\ref{fig:FigThermalFreeP}(a-ii) tells us that $T\ts{C}^{\text{eff}}/\Omega_1 > T\ts{H}^{\text{eff}}/\Omega_2$, and the statement \eqref{eq:free p necessary condition} implies that either work cannot be extracted or that the thermodynamic cycle is not closed.

Meanwhile, we can find a set of parameters that enables us to extract positive work from a closed QOE. 
This scenario is demonstrated in Fig.~\ref{fig:FigThermalFreeP}(b), where $T\ts{H}=4$ is chosen. 
Figure~\ref{fig:FigThermalFreeP}(b-i) shows that the extracted work is positive for all $\sigma$, which is consistent with the positive work condition examined in \ref{fig:FigThermalFreeP}(b-ii). 
This time, from Fig.~\ref{fig:FigThermalFreeP}(b-iii), there exists a value of $\sigma$ that satisfies the cyclicity condition while $\braket{W\ts{ext}}/\lambda^2 \sigma >0$. 
Moreover, Figs.~\ref{fig:FigThermalFreeP}(b-ii) and (b-iii) exemplify the statement \eqref{eq:free p necessary condition}: 
if positive work is extracted from a genuine QOE, then the effective temperatures satisfy $T\ts{C}^{\text{eff}}/\Omega_1 < 1/\ln[ (1-p)/p ] < T\ts{H}^{\text{eff}}/\Omega_2$, but the opposite is not always true.

\subsubsection{Case: adjusted $p$}

\begin{figure}[t]
\centering
\includegraphics[width=\linewidth]{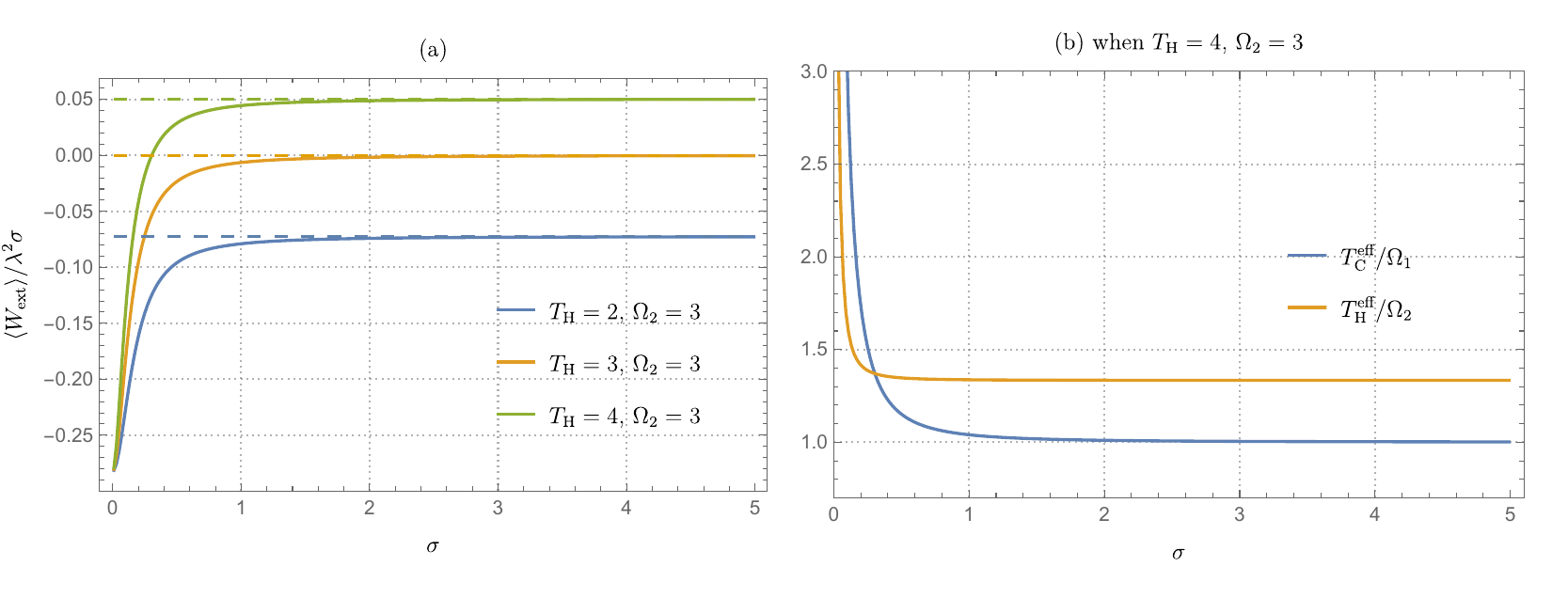}
\caption{(a) The extracted work $\braket{W\ts{ext}}/\lambda^2 \sigma$ as a function of the interaction duration $\sigma$ in units of $\Omega_1$. 
Here, $T\ts{C}=1$. 
The dashed lines are the work extracted \eqref{eq:work in detailed balanced} after the detailed balance condition is satisfied. 
(b) The effective temperature divided by the energy gap in the case of $T\ts{H}=4$ and $\Omega_2=3$ in (a). 
The positive work condition is satisfied when $T\ts{C}^{\text{eff}}/\Omega_1 < T\ts{H}^{\text{eff}}/\Omega_2$. 
Negative values mean that work is done \textit{on} the system (i.e., the UDW detector).}
\label{fig:WorkThermal}
\end{figure}

We now consider the QOE described in Sec.~\ref{subsubsec:when p is adjusted}. 
We first note that an inertial detector in the Minkowski vacuum (i.e., zero temperature: $T\ts{H}=T\ts{C}=0$) cannot extract work from vacuum fluctuations since the condition \eqref{eq:positive work condition wrt response} cannot be satisfied no matter what $\Omega$ and $\sigma$ are. 
We then consider  the KMS temperatures $T\ts{H}, T\ts{C} \neq 0$  
and compute the extracted work $\braket{W\ts{ext}}/\lambda^2 \sigma$. 
The interaction duration $\sigma$ is divided so that the perturbative analysis is valid---otherwise, the extracted work $\braket{W\ts{ext}}/\lambda^2$ monotonically increases with $\sigma$ without an upper bound. 
In Fig.~\ref{fig:WorkThermal}(a), we plot the extracted work $\braket{W\ts{ext}}/\lambda^2 \sigma$ as a function of $\sigma$ with various $T\ts{H}$. 
Here, quantities are in units of the energy gap $\Omega_1$ at the beginning of the cycle, and we fix $\Omega_2/\Omega_1=3$ and $T\ts{C}/\Omega_1=1$. 
We observe that the detector cannot extract positive work when the interaction duration is too short (i.e., $\sigma \Omega_1 \ll 1$), while $\braket{W\ts{ext}}/\lambda^2 \sigma$ asymptotes to a number at $\sigma \Omega_1 \gg 1$. 
This is the opposite behavior of the result shown in Fig.~\ref{fig:FigThermalFreeP}, and can be attributed to the difference between the positive work conditions given by \eqref{eq:free p positive work condition} and \eqref{eq:positive work condition wrt effective temp}. 
In the case of free $p$, as illustrated in Fig.~\ref{fig:FigThermalFreeP}(b), $T\ts{H}^{\text{eff}}/\Omega_2$ is compared to a constant value. 
In contrast, for Fig.~\ref{fig:WorkThermal}, $T\ts{H}^{\text{eff}}/\Omega_2$ is compared to another effective temperature $T\ts{C}^{\text{eff}}/\Omega_1$ as depicted in \ref{fig:WorkThermal}(b).

The asymptotic value for each curve is depicted by the dashed line in Fig.~\ref{fig:WorkThermal}(a), which is evaluated from \eqref{eq:work in detailed balanced}. 
This confirms that, at the long interaction duration, the detailed balance condition for the response function \eqref{eq:detailed balance for response} is satisfied. 
In addition, we observe that the work extracted when $(T\ts{H}/\Omega_1, \Omega_2/\Omega_1)=(2,3)$ and $(3,3)$ is always non-positive, whereas it is possible to extract positive work from the system with $(T\ts{H}/\Omega_1, \Omega_2/\Omega_1)=(4,3)$. 
This can be understood from the condition \eqref{eq:positive work condition wrt temp}; when the detailed balance condition \eqref{eq:detailed balance for response} is satisfied ($\sigma \Omega_1 \gg 1$ in this case), the detector can extract positive work when the temperature of the baths and the energy gap obey \eqref{eq:positive work condition wrt temp}.

\subsection{UDW detector in circular motion}\label{subsec:circular}

The Unruh QOE utilizes the thermality caused by the Unruh effect at the temperatures $T\ts{H}=a\ts{H}/2\pi$ and $T\ts{C}=a\ts{C}/2\pi$ with $a\ts{H}>a\ts{C}$ in the protocol described in Sec.~\ref{sec: QOE} \cite{UnruhOttoEngine}. 
However, as we pointed out in the Introduction, a linearly accelerating UDW detector is not ideal for extracting thermodynamic work since it requires a tremendous amount of acceleration $a$ as well as a huge space to let the detector travel long distances. 
Instead, we wish to employ a UDW detector in circular motion, which not only allows us to confine the detector in a compact space, but also less work is required to acquire some temperatures compared to the linear acceleration case. 
Moreover, the Wightman function along a circulating detector's trajectory does not satisfy the KMS condition. 
Our general expression for the extracted work \eqref{eq:free p work extracted} and \eqref{eq:extracted work wrt response} as well as the positive work condition 
\eqref{eq:free p positive work condition} and \eqref{eq:positive work condition wrt effective temp} can still be applied to such a scenario.


Consider the circular trajectory of the UDW detector, which is given by 
\begin{align}
    \sx(\tau)
    &=
        \{ t=\gamma \tau, x=R \cos (\omega \gamma \tau)\,,y=R\sin (\omega \gamma \tau), z=0  \}\,. \label{eq:circular trajectory}
\end{align}
Here, $R(>0)$ is the radius of the circle, $\omega$ is the angular velocity of the detector, and $\gamma = 1/\sqrt{1-R^2 \omega^2}$. 
One can introduce the proper acceleration of the detector, whose magnitude $a$ is given by $a=R \omega^2 \gamma^2$, as well as the speed of the detector $v=R \omega(\leq 1)$. 
Using these relations, one can write $\gamma, \omega, v$ in terms of $a$ and $R$ as follows. 
\begin{subequations}
    \begin{align}
    \omega 
    &= 
        \sqrt{ \dfrac{a}{ (1+ a R ) R } }\,, \\
    \gamma
    &=
        \sqrt{ 1 + a R }\,, \\
    v 
    &= 
        \sqrt{ \dfrac{ a R }{ 1 + a R } }\,.
\end{align}
\end{subequations}

In order to calculate the response function \eqref{eq:response function}, one needs to have an expression of the pullback of the Wightman function along the trajectory \eqref{eq:circular trajectory}. 
Consider a minimally coupled, massless quantum scalar field in $(3+1)$-dimensional Minkowski spacetime. 
The Wightman function in the Minkowski vacuum is given by \eqref{eq:Wightman Minkowski}. 
Inserting \eqref{eq:circular trajectory}, we obtain the pullback of the Wightman function: 
\begin{align}
    &W(\sx(\tau), \sx'(\tau'))
    =
        -\dfrac{1}{4\pi^2}
        \dfrac{1}{\gamma^2(\Delta \tau - \ii\epsilon)^2 - 4R^2\sin^2(\omega\gamma \Delta \tau/2)} \,,
\end{align}
where $\Delta \tau \coloneqq \tau-\tau'$. 
Since the Wightman function is time-translation invariant, the response functions do not depend on the switching time.


\begin{figure*}[t]
\centering
\includegraphics[width=\linewidth]{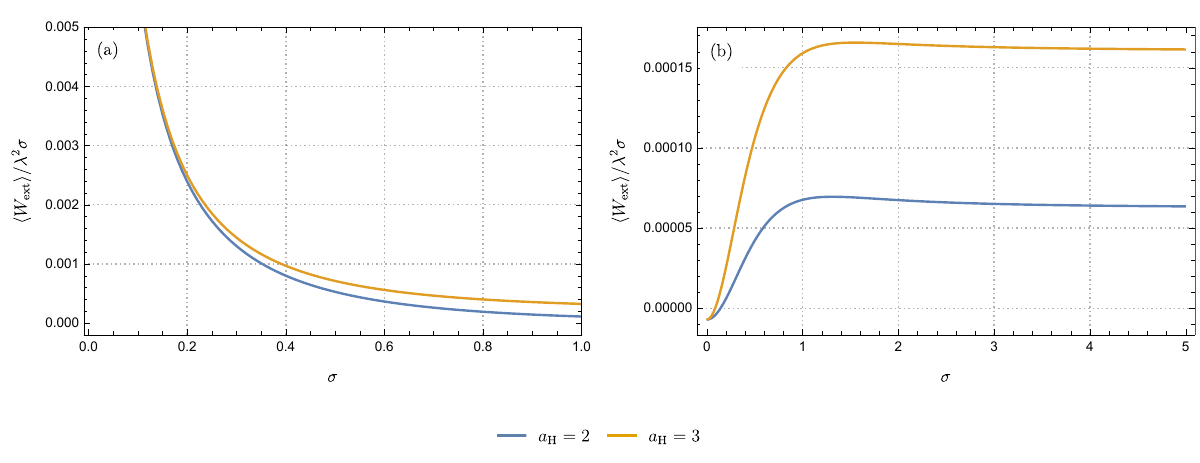}
\caption{The extracted work $\braket{W\ts{ext}}/\lambda^2 \sigma$ by a detector in circular motion as a function of the interaction duration $\sigma$ in units of $\Omega_1$. 
Here, $\Omega_2=1.01$ and $a\ts{C}=1$. 
(a) The case where $p$ is adjustable and $p=1/20$, and (b) the case where $p$ is tuned to complete the cycle. 
Each figure contains $a\ts{H}=2$ and $a\ts{H}=3$. 
}
\label{fig:FigCircvsSigma}
\end{figure*}

\subsubsection{Extracted work}

From the Wightman function above, we can numerically calculate the response functions and the work extracted for the detector in circular motion. 
In Fig.~\ref{fig:FigCircvsSigma}, we plot $\braket{W\ts{ext}}/\lambda^2 \sigma$ as a function of $\sigma$ in units of the energy gap $\Omega_1$ in step-1. 
Figures~\ref{fig:FigCircvsSigma}(a) and \ref{fig:FigCircvsSigma}(b) are respectively the cases where $p$ is freely chosen and tuned. 
In both figures, we set $\Omega_2/\Omega_1=1.01, R\Omega_1=1$, and $a\ts{C}/\Omega_1=1$, and each figure includes the curves for $a\ts{H}/\Omega_1=2$ and 3.

As in the case of the inertial detector in a thermal bath, QOE with a free $p$ (here $p=1/20$) extracts more work when the interaction duration is short [Fig.~\ref{fig:FigCircvsSigma}(a)], whereas if $p$ is tuned to close the cycle, the opposite behavior holds [Fig.~\ref{fig:FigCircvsSigma}(b)]. 
However, we find in the latter case that, unlike in Sec.~\ref{subsec:KMS}, the extracted work does not reach a maximum at large $\sigma$. 
Instead, there exists a maximum at $\sigma \Omega_1 \approx 1$. 
Nevertheless, we confirm that this behavior is contingent on the choice of parameters.

\begin{figure}[t]
\centering
\includegraphics[width=\linewidth]{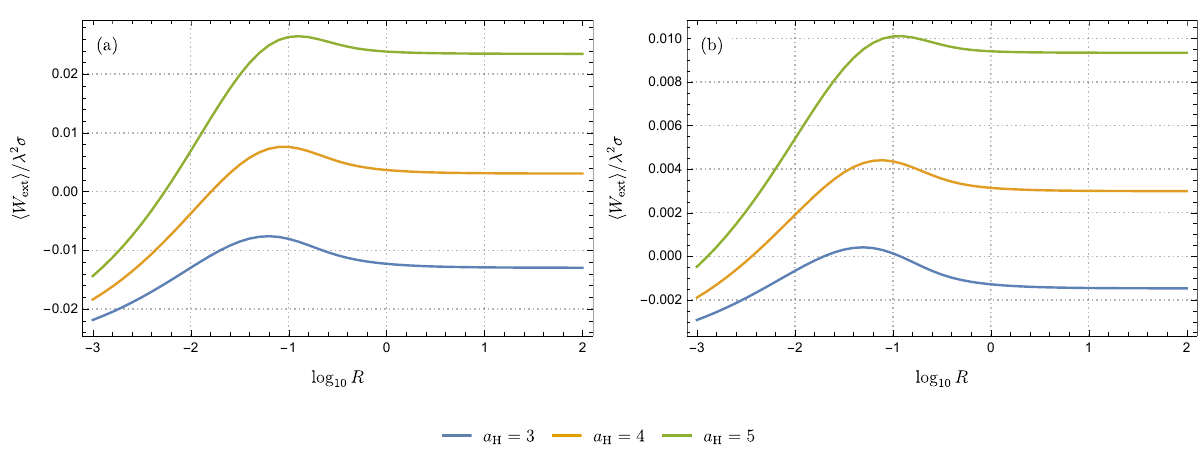}
\caption{Work extracted by a circulating UDW detector as a function of the radius $\log_{10}R$ when (a) $p$ is adjustable (chosen to be $p=1/20$), and (b) $p$ is tuned to complete the cycle. 
Here, $\Omega_2=2, \sigma=1$, and $a\ts{C}=1$ in units of $\Omega_1$. 
Each figure contains three cases of $a\ts{H}$: $a\ts{H}=3,4$, and 5. Note that
negative values correspond to  work done \textit{on} the system (i.e., the UDW detector).}
\label{fig:FigWextCircvslogR}
\end{figure}

We then plot $\braket{W\ts{ext}}/\lambda^2 \sigma$ as a function of the radius of the circle, $\log_{10} R$, in units of $\Omega_1$ in Fig.~\ref{fig:FigWextCircvslogR}. 
Here, we set $\Omega_2=2, \sigma=1, a\ts{C}=1$, and each figure includes curves for $a\ts{H}=3, 4$, and 5. 
As before, Fig.~\ref{fig:FigWextCircvslogR}(a) corresponds to QOE with $p=1/20$, while \ref{fig:FigWextCircvslogR}(b) depicts the result with a tuned $p$. 
Both figures exhibit a similar pattern; a detector orbiting with a small radius $R$ has difficulty extracting work, and $\braket{W\ts{ext}}/\lambda^2 \sigma$ is maximized as $R$ increases, eventually asymptoting to some value at $R \to \infty$. 
Thus, the radius $R$ of the orbit must be optimized to extract the maximum amount of work.

\subsubsection{Fixed acceleration}

Although the analytic expression for the effective temperature for a circulating detector, $T\ts{circ}$, cannot be obtained in general, one can impose an additional assumption to obtain the exact form of $T\ts{circ}$. 
Suppose that the detector is circulating at high speed $v\approx 1$ (i.e., ultrarelativistic regime), which can be realized by taking a large radius $R$. 
The effective temperature in this case is known to be \cite{CircularTemperaturesUnruh, Unruh.acceleration.rad.electron.1998.arxiv} 
\begin{align}
    T\ts{circ}
    &=
        \dfrac{ |\Omega| }{ \ln \kagikako{ 1 + \dfrac{4 \sqrt{3} |\Omega| }{ a } \exp \kako{ \dfrac{2 \sqrt{3} |\Omega| }{ a } } } } 
        \in ( a/4\sqrt{3}, a/2\sqrt{3} )\,. \label{eq:ultrarelativistic temp circular}
\end{align}
While the effective temperature $T\ts{circ}$ increases with $a$, it is also possible to enhance the temperature by enlarging the energy gap $\Omega$ with fixed acceleration. 
That is, for a given $a$, $T\ts{circ}(\Omega_2) > T\ts{circ}(\Omega_1)$ if $\Omega_2 > \Omega_1$. 
We then ask if the circulating detector at high speed with a fixed acceleration $a$ can extract positive work by manipulating the effective temperature by adjusting the energy gap.

The answer to this inquiry is negative. 
To see this, it is easy to check that, for a given $a>0$, $\Omega_2 > \Omega_1 \Rightarrow T\ts{circ}(\Omega_2) /\Omega_2 < T\ts{circ}(\Omega_1) /\Omega_1$, which violates the necessary condition for positive work, \eqref{eq:free p positive work condition} and \eqref{eq:positive work condition wrt effective temp}. 
Thus, we find that, for any $p\in (0,1)$, it is impossible to extract positive work from a UDW detector in a circular motion at $v\approx 1$ by fixing $a$. 
This is one example of the feature of quantum heat engines: $T\ts{C}< T\ts{H}$ does not necessarily guarantee the extraction of positive work in quantum heat engines, but rather, one should consider $T\ts{C}/\Omega_1 < T\ts{H}/\Omega_2$, where $\Omega_1 < \Omega_2$.

\section{Conclusion}
\label{sec: conclusion}

Inspired by the Unruh quantum Otto engines, we considered the quantum Otto engine (QOE) in a general setting utilizing an Unruh-DeWitt (UDW) detector interacting with a quantum scalar field. 
While the original QOE considered a qubit thermalized in thermal baths, we assumed (i) the field is in a quasi-free state, not necessarily in a Kubo-Martin-Schwinger (KMS) thermal state, and (ii) the detector follows an arbitrary trajectory in curved spacetime. 
By carrying out a perturbative method, we managed to derive the extracted work \eqref{eq:free p work extracted} and \eqref{eq:extracted work wrt response} in terms of the response function of the UDW detector. 
In addition, we found that the condition for extracting positive work can be expressed in terms of the effective temperature perceived by the detector, \eqref{eq:free p positive work condition} and \eqref{eq:positive work condition wrt effective temp}. 
These formulae are applicable to various scenarios outside (thermal) equilibrium. 
As a particular case, if the field is in the KMS state and the detector with a rapidly decreasing switching function interacts with it for a long time, the response function is known to obey the detailed balance condition, which makes the effective temperature identical to the KMS temperature. 
The condition \eqref{eq:positive work condition wrt effective temp} then becomes the positive work condition \eqref{eq:positive work condition wrt temp} originally found by \cite{Feldmann.QOE.2000, Kieu.secondlaw.demon.otto}.

Using the formula for the extracted work, we explored two concrete examples: an inertial detector in the KMS state and a circulating detector in the Minkowski vacuum.

In the scenario of an inertial detector in thermal baths, we numerically examined the effect of the interaction duration. 
The relationship between the extracted work and the interaction duration is drastically changed by the choice of the detector's initial state. 
Nevertheless, the extracted work asymptotes to the value that corresponds to thermal equilibrium as the interaction duration increases.

Another example involves a circulating detector in the Minkowski vacuum. 
The motivation for this scenario comes from the previously proposed Unruh QOE \cite{UnruhOttoEngine}, in which the thermal baths are generated by the Unruh effect. 
However, utilizing a linearly accelerating UDW detector in the Unruh QOE is not practical for several reasons such as the work necessary for acceleration and the requirement for a large space. 
The circulating detector, on the other hand, circumvents these issues. 
Moreover, the QOE with a circulating detector can be adequately examined only when our general framework is employed; 
The Wightman function pulled back along the circular trajectory does not satisfy the KMS condition, which suggests that the use of the effective temperature is necessary.

The work extracted by the circulating detector showed behavior similar to that of the inertial detector in a thermal bath. 
However, we found that there exist optimal values for the interaction duration and radius of the orbit to extract maximum work. 
Furthermore, we showed that the detector is unable to extract work with a fixed acceleration due to the violation of the positive work condition.

 An interesting extension to our work would involve exploring quantum heat engines using non quasi-free states.
In particular, coherent and squeezed states---which are not quasi-free---are known to be particular interest in a variety of settings. 
It would be interesting to see how a non-vanishing one-point correlator contributes to the cycle.

\section*{Acknowledgments}
This work was supported in part by the Natural Sciences and Engineering Research Council of Canada.
KGY is thankful to Dr. Jorma Louko for the insightful discussion.


\bibliography{ref}
\bibliographystyle{JHEP}

\end{document}